\documentclass[useAMS,usenatbib]{mn2e}
\usepackage{times}

\usepackage{graphics,epsfig}

\title[BTF Relation for Extremely Low Mass Galaxies]{Baryonic Tully-Fisher Relation for Extremely Low Mass Galaxies}
\author[Begum et al.]
{
Ayesha Begum$^{1}$\thanks{E-mail:ayesha@ast.cam.ac.uk},
Jayaram N. Chengalur$^{2}$, 
I. D. Karachentsev$^{3}$ and
M. E. Sharina$^{3}$
\\
\\
$^{1}$Institute of Astronomy, University of Cambridge, Madingley Road, Cambridge, CB3 0HA, UK\\
$^{2}$National Centre for Radio Astrophysics, Post Bag 3, Ganeshkhind, Pune 411 007, India\\
$^{3}$Special Astrophysical Observatory, Nizhnii Arkhys 369167, Russia\\
}
\begin{document}

\date{}


\maketitle

\label{firstpage}

\begin{abstract}
  We study Tully Fisher relations for a sample that combines extremely faint (M$_{\rm B} < -14.$) galaxies along with bright (i.e. $\sim L_{*}$) galaxies. Accurate ($\sim 10\%$) distances, I band photometry, and B-V colors are known for the majority of the galaxies in our sample. The faint galaxies are drawn from the Faint Irregular Galaxy GMRT survey (FIGGS), and we have HI rotation velocities derived from aperture synthesis observations for all of them. For the faint galaxies, we find that even though the median HI and stellar masses are  comparable, the HI mass correlates significantly better with the circular velocity indicators than the stellar mass. We also find  that the velocity width at the 20\% level (W$_{20}$) correlates better with mass than the rotation velocity, although the difference is not statistically significant. The faint galaxies lie systematically below the I band TF relation defined by bright galaxies, and also show significantly more intrinsic scatter. This implies that the integrated star formation in these galaxies has been both less efficient and also less regulated than in large galaxies. We estimate the intrinsic scatter of the faint galaxies about the I band TF to be $\sim 1.6$~mag. We find that while the faint end deviation is greatly reduced in Baryonic Tully-Fisher (BTF) relations, the existence of a break at the faint end of the BTF is subject to systematics such as the assumed stellar mass to light ratio. If we {\it assume} that there is an intrinsic BTF and try to determine the baryonic mass by searching for prescriptions that lead to the tightest BTF, we find that scaling the HI mass leads to a much more significant tightening than scaling the stellar mass to light ratio. The most significant tightening that we find however, is if we scale the entire baryonic mass of the faint (but not the bright) galaxies. Such a scenario would be consistent with models where dwarf (but not large) galaxies have a large fraction of dark or ``missing'' baryons. In all cases however the minimum in the $\chi^2$ curve is quite broad and the corresponding parameters are poorly constrained.
\end{abstract}


\section{Introduction}
\label{sec:intro}

    One of the stringent tests of galaxy formation models is their ability to reproduce the tightness of the Tully-Fisher (TF) relation, i.e. the fact that the rotation velocity and absolute magnitude of bright spiral galaxies are tightly correlated. For such galaxies, the intrinsic scatter in the I band TF has been estimated to be as small as 0.2 mag (e.g. \cite{sakai00}), while \cite{verheijen01} argues that there is probably no intrinsic scatter in the K$^{'}$ band TF relation. This implies that the dark and visible (i.e. stellar) matter content of galaxies are tightly correlated, or, if one assumes that the visible matter traces the bulk of the baryons, that the non baryonic and baryonic content of galaxies are tightly correlated. Theoretically, one would not expect this correlation to hold all the way down to dwarf galaxy masses. Reheating prior to the epoch of reionization would lead to small halos having a reduced baryon fraction (e.g. \cite{gnedin00}), and small dark matter halos are also expected to easily lose baryons because of energy input from  the first burst of star formation  and supernova winds (e.g. \citealp{ts07,dw03,ef00}). Large galaxies on the other hand are more able to keep their original share of baryons.  Consistent with this expectation, observations show that dwarf galaxies with  rotation velocities less than  $\sim$ 90 kms$^{-1}$ lie below the  TF relation defined by brighter galaxies(\citealp{mg00,bc04,mg05}), i.e. they are under luminous for the velocity width. On the other hand, \cite{geha06} using an independent SDSS based sample of dwarf galaxies did not find any break in the I band TF relation, although, in agreement with earlier studies they find an increased scatter about the TF relation at low luminosities. Further, as first noted by \cite{mg00} the ``break" seen at the faint end of the  B band TF relation is removed  if one works with the total baryonic mass (i.e. the sum of stellar and gas)  instead of the luminosity.

    Accurate extensions of the TF relation to faint dwarf galaxies poses a number of practical problems. Firstly, dwarf galaxies are largely a field population, and very few of them have accurately known distances. Secondly, being faint, and largely of low surface brightness, accurate photometry is difficult. Finally, in contrast to the situation for bright spirals, the peak rotational velocities of faint (M$_{\rm B} > -14$) dwarf galaxies is generally comparable in magnitude to the velocity dispersion of the gas (e.g. \citealp{b1,bc04}). In such a situation it is unclear if the velocity width (e.g. at the 20\% level, W$_{\rm 20}$, which is the commonly used indicator of the circular velocity in TF relations), is a good tracer of the circular velocity. 

In this paper we discuss the TF relation using a data set which address all of the above issues. Accurate ($\sim 10\%$) TRGB distances are known to most of the galaxies in our sample, and we also use accurate I band magnitudes. Finally all of the galaxies have been observed in HI using the Giant Meterwave Radio Telescope (GMRT) as part of the FIGGS survey (see \cite{begum08}). Pressure corrected rotation curves are available for all of the galaxies; these can be compared with W$_{\rm 20}$ to see which is the better observable to use for TF studies. As a comparison sample we use the sample of bright galaxies for which Cephied distances were measured as part of the HST key project on the extra galactic distance scale (\cite{sakai00}).

\section[]{The Dwarf Galaxy Sample }
\label{sec:obs}

\begin{table*}
\caption{FIGGS data}
\label{tab:data}
\begin{tabular}{|l|c|c|c|c|c|c|c|c|c|c|c|c|}
\hline
Galaxy &
V$_{rot}$ &
  \multicolumn{1}{c|}{V$_{der}$} &
  \multicolumn{1}{c|}{V$_{err}$} &
  \multicolumn{1}{c|}{M$_B$} &
  \multicolumn{1}{c|}{M$_I$} &
  \multicolumn{1}{c|}{M$_{HI}$} &
  \multicolumn{1}{c|}{L$_B$} &
  \multicolumn{1}{c|}{$\frac{D_{HI}}{D_{Ho}}$} &
  \multicolumn{1}{c|}{B-V} &
  \multicolumn{1}{c|}{W$_{20}$}& 
D &
Method\\
& (kms$^{-1}$)&(kms$^{-1}$)&(kms$^{-1}$)&(mag)&(mag)&(10$^6 M_\odot$)&(10$^6$ L$_\odot$)&&(mag)&(kms$^{-1}$) &(Mpc)&\\
\hline
  UGC 685      & 51.67 & 45.84 & 2.75 & $-$14.31 & $-$15.07 & 56.15  & 82.41 &  1.6 & 0.52 & 74.0 & 4.5 &rgb\\
  KKH 6        & 27.98 & 19.93 & 3.66 & $-$12.42 & $-$13.53 & 10.18  & 14.45 &  2.9 & 0.43 & 40.0 & 3.73&rgb\\
  KK 14$^a$    & 21.32 & 19.27 & 1.52 & $-$12.13 & $-$12.84 & 21.92  & 11.06 &  1.5 & 0.42 & 36.5 & 7.2 & grp\\
  KKH 11$^a$   & 56.29 & 52.39 & 3.14 & $-$13.35 & $-$14.06 & 52.87  & 34.04 &  4.2 & 0.40 & 106.0& 3.0 & grp\\
  KK 41        & 32.3  & 21.91 & 1.97 & $-$14.06 & $-$15.27 & 67.19  & 65.46 &  3.3 & 0.63 & 55.0 & 3.9 &rgb\\
  UGCA 92      & 37.08 & 31.00 & 3.00 & $-$15.65 & $-$15.71 & 156.05 & 283.13&  4.5 & 0.15 & 70.0 & 3.01&rgb\\
  KK 44        & 20.00 & 7.00  & 2.10 & $-$11.85 & $-$13.03 & 12.0   & 8.57  &  2.3 & 0.58 & 32.6 & 3.34&rgb\\
  KKH 34       & 20.11 & 10.50 & 1.38 & $-$12.30 & $-$13.23 & 10.44  & 12.94 &  2.6 & 0.28 & 25.0 & 4.6 &rgb\\
  UGC 3755     & 20.19 & 13.05 & 2.73 & $-$14.90 & $-$16.00 & 73.99  & 141.90& 1.7  & 0.55 & 49.0 & 6.96&rgb\\
  DDO 43       & 36.04 & 30.72 & 2.16 & $-$14.75 & $-$15.49 & 203.02 & 123.59&  2.8 & 0.45 & 50.0 & 7.8 &rgb\\
  KK 65        & 18.49 & 16.71 & 2.90 & $-$14.29 & $-$15.57 & 38.85  & 91.20 &  2.3 & 0.54 & 50.2 & 7.62&rgb\\
  UGC 4115     & 58.54 & 56.08 & 2.58 & $-$14.27 & $-$15.95 & 285.52 & 79.43 & 4.0  & 0.47 & 105.0&7.5  &rgb\\
  UGC 5456     & 39.75 & 37.55 & 5.29 & $-$15.08 & $-$16.23 & 58.95  & 167.49& 1.5  & 0.39 & 88.0 & 5.6 &rgb\\
  UGC 6145$^a$ & 40.90 & 36.94 & 2.90 & $-$13.14 & $-$13.85 & 27.02  & 28.05 &  1.6 & 0.40 & 52.0 & 7.4 &h\\
  NGC 3741     & 46.70 & 46.70 & 2.50 & $-$13.13 & $-$13.94 & 157.99 & 27.79 &  8.8 & 0.37 & 101.0& 3.0 &rgb\\
  UGC 7242     & 39.90 & 35.11 & 3.03 & $-$14.06 & $-$14.77 & 45.75  & 65.46 &  2.1 & 0.40 & 74.0 & 5.4 &rgb\\
  KK 144       & 27.82 & 22.38 & 3.95 & $-$12.59 & $-$14.32 & 81.14  & 16.90 & 3.1  & 0.38 & 49.0 & 6.3 &h\\
  DDO 125      & 22.92 & 14.61 & 2.74 & $-$14.16 & $-$15.24 & 31.87  & 71.77 &  1.7 & 0.51 & 42.0 & 2.5 &rgb \\
  UGC 8055     & 65.72 & 60.25 & 2.72 & $-$15.49 & $-$16.46 & 782.63 & 244.34&  3.0 & 0.37 & 95.5 & 17.4&tf\\
  UGC 8215     & 19.95 & 12.37 & 4.00 & $-$12.26 & $-$13.30 & 21.41  & 12.47 &  3.5 & 0.38 & 31.0 & 4.5 &rgb\\
  UGC 8508     & 40.10 & 31.17 & 2.30 & $-$12.98 & $-$13.99 & 29.07  & 24.21 &  3.3 & 0.46 & 64.5 & 2.6 &rgb\\
  DDO 181      & 29.92 & 22.17 & 4.00 & $-$13.03 & $-$13.90 & 27.55  & 25.35 &  3.3 & 0.46 & 49.0 & 3.1 &rgb\\
  DDO 183      & 24.66 & 21.35 & 1.85 & $-$13.17 & $-$13.97 & 25.90  & 28.84 &  2.7 & 0.42 & 48.0 & 3.24&rgb\\
  UGC 8833     & 26.68 & 20.65 & 3.01 & $-$12.42 & $-$13.27 & 15.16  & 14.45 &  2.3 & 0.42 & 40.0 & 3.2 &rgb\\
  P51659$^a$   & 30.50 & 22.00 & 2.58 & $-$11.83 & $-$12.54 & 52.99  & 8.39  &  2.7 & 0.40 & 55.5 & 3.6 &rgb\\
  KK 246       & 35.18 & 29.93 & 4.00 & $-$13.69 & $-$14.87 & 63.39  & 46.47 &  2.9 & 0.58 & 80.0 & 7.83&rgb\\
  KK 250$^b$   & 50.00 & 50.00 & 3.00 & $-$14.54 & $-$16.14 & 121.0  & 120.26& 3.2  & 0.91 & 110.0& 5.6 &  grp\\
  KK 251$^b$   & 38.00 & 34.00 & 4.00 & $-$13.72 & $-$14.32 & 78.0   & 44.05 & 2.6  & 0.37 & 79.0 & 5.6 & grp\\
  DDO 210$^c$  & 17.00 & 8.00  & 2.00 & $-$11.09 & $-$11.90 & 2.80   & 4.30  &  1.3 & 0.24 & 29.1 & 1.0 &rgb\\
\hline
\end{tabular}
$^a$ I band magnitude computed using the relation (B-V)=0.85*(V-I) (\cite{makarova99}) assuming B-V=0.4 and B magnitudes from \cite{kk04}.
$^b$ I band magnitude from \cite{bc04}.
$^c$ I band magnitude from \cite{lee99}.
\end{table*}

The dwarf galaxy sample is a part of an HI imaging study of faint dwarf galaxies with the GMRT $-$ the Faint Irregular Galaxies GMRT Survey (FIGGS; \cite{begum08}). The FIGGS sample is a systematically selected subsample of the Karachentsev et al.(2004) catalog of galaxies within 10 Mpc (see \cite{begum08} for details on the sample selection and observations). The FIGGS galaxies represent the extreme low-mass end of the dIrr population, with a median ${\rm{M_B}} \sim -13$ and a median HI mass $\sim 3 \times 10^7$~M$_\odot$.  Our GMRT observations show that ,contrary to the general belief that extremely faint dIrr galaxies have ``chaotic" velocity fields (\cite{lo93}), most of them have coherent large scale velocity gradients (see e.g. \cite{bc06}). For many galaxies, these 
large scale gradients are consistent with the systematic rotation. 
Rotation curves were derived using a tilted ring model, details can be found in Begum et al.(2008, in preparation). For the current study, we have considered only those galaxies from the FIGGS sample 
which are well resolved (i.e.are more than 6 independent beams across) and for 
which tilted ring analysis gave sensible fits to the data. 
Examples can be seen in, for e.g. \cite{b1,bc04,bc05,bc06}.
The galaxies from the FIGGS sample selected for the current study are given in Table~\ref{tab:data}. The columns are as follows:
Column (1) the name of the galaxy. 
Column (2) V$_{rot}$, the rotation velocity  at the last measured point of the rotation curve, corrected for the pressure support,(see e.g. \citealp{b1,bc04}).
Column(3) V$_{der}$, the derived rotation velocity at the last measured point of the rotation curve, 
Column(4) V$_{err}$, the error on the derived velocity, 
Column(5) the absolute B magnitude, 
Column(6) the absolute I magnitude. 
Column (7)the HI mass derived from the GMRT data,
Column(8) the luminosity in B band, 
Column(9) the extent of the HI disk at a level of N$_{HI}=1\times 10^{19}$cm$^{-2}$, normalized to the Holmberg radius of the galaxy. 
Column(10) the B-V colour, 
Column (11)the velocity width of the global HI profile at 20\% level, derived from the GMRT data, 
Column (12)the distance to the galaxy from  \cite{kk04} and 
Column(13) the method used to derive the distance e.g.  the tip 
of the red giant branch (rgb), Tully-Fisher relation (tf),  the Hubble relation (h) and the member ship of the group (grp). 22 of the 29 galaxies in our sample have independent distances measured using the TRGB. In our analysis below, where appropriate, we treat the galaxies with independent distances separately from those which do not.

The sources of the absolute magnitudes of our sample galaxies are listed in the table. The photometry for those galaxies for which there is no separate notation is from an HST (WFPC2 and ACS) survey of Local Volume dwarf galaxies. The photometry of galaxies with large angular diameters was done using SDSS images.  The HST photometry was done using the recipes and  equations given in  Holtzman et al.(1995) and Sirianni et al.(2005). Details on both the HST and SDSS based photometry can be found in (\cite{sme07}). Photometry for the few galaxies in our sample that are not also in \cite{sme07} was done using SDSS images and an identical procedure to that followed by \cite{sme07}. Transformation to the standard  Johnson-Cousins system was done using the empirical color transformations given by Jordi et al.(2006). Note that these magnitudes are those measured within a limiting diameter (tabulated in \cite{sme07}). In principle one
could instead use extrapolated total magnitudes derived from 
assuming an exponential stellar disk. However, in practice, the surface brightness distribution is poorly fit by an exponential disk, and there is substantial uncertainty in the scale length and location of center of the stellar disk.Consequently, the errors in the extrapolated total magnitudes are large, and, in fact, will agree within the error bars with those listed in Table~\ref{tab:data}. We hence use the magnitudes as tabulated by \cite{sme07}.

 As a comparison bright galaxy sample we use a sample of 21 galaxies with 
Cepheid distances taken from \cite{sakai00}. 
Rotation curves are available  in literature for only few galaxies in this sample, hence,
we use the inclination corrected velocity widths at 20\% level of the peak emission ( W$_{20}$), given in  \cite{sakai00}. The velocity widths were corrected for the line broadening due to the turbulent
velocity dispersion using the prescription  given in Verheijen \& Sancisi (2001).
Note that  although there exist several intermediate mass dwarf galaxies
samples (e.g. Schombert et al. 1997) that could have  been used as
templates, we have restricted our analysis here to galaxy samples with
accurate, Hubble flow independent distance estimates.

\section[]{Results and Discussion}
\label{sec:res}

    Table~\ref{tab:corrcoef} lists the correlation coefficients between different indicators of the mass and the circular velocity.
We use either V$_{\rm rot}$ or W$_{20}$ after correction for the line broadening due to the turbulent velocity dispersion using the prescription  given in Verheijen \& Sancisi (2001).
The stellar mass, $M_{*}$ was computed using the stellar mass to light ratio given by \cite{bell03} for the case of a ``diet" Salpeter IMF. Two estimates of the total baryonic mass were used; the first (M$_{\rm bar}$) was computed using the sum of the stellar mass and 1.33 times the HI mass to correct for primordial He (Ott et al.(2001);
Cote et al. (2000)).
No correction for the molecular gas was made, which is probably 
justifiable for the FIGGS sample, since dwarf galaxies do not seem to have a substantial reservoir of molecular gas (\cite{taylor98}). While the molecular gas probably makes a substantial contribution to the total gas mass in the case of the bright spirals in the Sakai et al. (2000) sample, the baryonic mass of these systems is dominated by the stellar contribution, and the error we make in ignoring the molecular gas is likely to be subsumed by the uncertainty in the stellar mass to light ratio. The second estimate of the baryonic mass (M$_{\rm bar_1}$) uses the same estimate of the total gas mass as before, but the stellar mass was computed from \cite{bell03} for a Bottema IMF. For a given B-V color, the mass to light ratio for the Salpeter ``lite'' IMF and Bottema IMF differ only by a multiplicative constant, so the correlation between the stellar mass computed using a Bottema IMF and the circular velocity indicators are identical to that computed using the Salpeter ``lite'' IMF. These two IMFs were chosen as being representative of two extremes of the stellar mass to light ratio(\cite{bell03}). The correlation coefficient error bars were computed using bootstrap resampling. The listed values are for the entire FIGGS sample; if one uses only the subsample of 22 galaxies with TRGB distances, the computed correlation coefficients agree within the error bars with the tabulated values. For the galaxies in the FIGGS sample alone,  a few interesting points emerge. The first is that even though the HI and stellar masses of these galaxies are comparable (the median ratio of the HI to stellar mass is $\sim 1.25$) the HI mass is significantly better correlated with the circular velocity  indicators than the stellar mass. In fact, the HI mass correlates with the circular velocity indicators at least as well the baryonic mass estimates. The second is that the error bars in the correlation coefficients are significantly larger for the FIGGS sample than for the combined HST+FIGGS sample. This is partly because of the substantially larger baseline of the combined sample, but as as shown below, it is also partly because of genuinely increased scatter in the low mass range. Finally, W$_{\rm 20}$ correlates better with the mass indicators than V$_{rot}$, although the correlation coefficients overlap within the error bars.

\begin{figure*}
\psfig{file=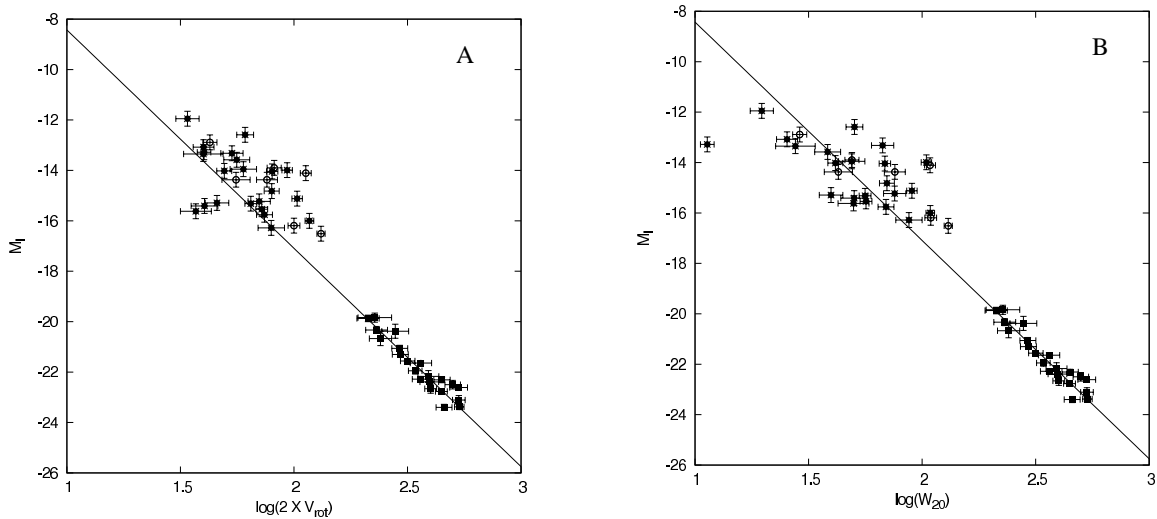,width=6.0truein}
\caption{[A]~The absolute I band magnitudes plotted as a function of the log of the rotation velocity V$_{\rm rot}$ for the FIGGS sample and W$_{20}$ (inclination and turbulence corrected) for galaxies from  the Sakai et al.(2000) sample. [B]~Same as [A] except that the abscissa is log(W$_{20}$) for both the FIGGS and Sakai et al. (2000)
samples. As can be seen, relative to the TF relation for bright galaxies, the FIGGS galaxies are under-luminous for their velocity width.
}
\label{fig:itf}
\end{figure*}

  The best fit regression of log(M$_{\rm I}$)  on log(W$_{\rm 20})$ computed using a bivariate least squares fit for the data from Sakai et al. (2000) gives log(M$_{\rm I}$) = -8.66 (0.67) log(W$_{\rm 20}) + 0.231(1.7)$ and is shown in Figure~\ref{fig:itf} as a solid line.
The error bars account for errors in both the distance measurement and the photometry. The two different panels show where the FIGGS galaxies lie with respect to this line, in Panel~[A] W$_{20}$ is used as the estimator of the circular velocity and in Panel~[B] one uses V$_{\rm rot}$. As can be seen, relative to the TF relation for bright galaxies, the FIGGS galaxies are under-luminous for their velocity width. However, the offset is significantly larger when one uses V$_{\rm rot}$ ($0.93 \pm 0.06$ mag) than when one uses W$_{\rm 20}$ ($0.31 \pm 0.06$) mag. If one does a regression fit using W$_{\rm 20}$
to the entire FIGGS+HST sample, the best fit relation is 
log(M$_{\rm I}$) = -8.77 (0.1) log(W$_{\rm 20}) + 0.741(0.3)$. The scatter of the FIGGS galaxies about this relation is $\sim 1.6$mag,
while that of the HST sample galaxies is $\sim 0.36$mag. The scatter for the FIGGS galaxies is much larger than the typical estimated total measurement error of $\sim 0.37$~mag, (from the errors in the distance measurement, the photometry and the error in the velocity width scaled by the TF slope) and indicates that essentially all of the observed scatter at the faint end is intrinsic.

\begin{table}
\caption{Correlation Coefficients}
\label{tab:corrcoef}
\begin{tabular}{lccc}
\hline
                    &\multicolumn{2}{c}{FIGGS}    & FIGGS+HST   \\
                    &log(V$_{\rm rot})$& log(W$_{\rm 20}$) &log(W$_{\rm 20}$) \\
\hline
\hline
log(M$_{\rm I}$)     &-0.505(0.15) &-0.629(0.10)  &-0.931(0.01) \\
log(M$_{*}$)         & 0.459(0.16) & 0.596(0.11)  & 0.928(0.02) \\
log(M$_{\rm HI}$)    & 0.683(0.10) & 0.733(0.07)  & 0.910(0.01) \\
log(M$_{\rm bar}$)   & 0.631(0.12) & 0.732(0.08)  & 0.942(0.01) \\
log(M$_{\rm bar_1}$) & 0.665(0.11) & 0.741(0.07)  & 0.943(0.01) \\
\hline
\end{tabular}
\end{table}

The systematic deviation from and the increased scatter around, the TF relation set by bright galaxies, implies that dwarf galaxies have been relatively less efficient at forming stars, and that there has also be considerably more stochasticity in their integrated star formation efficiencies. \cite{lee07} find that dwarf galaxies (i.e with M$_{\rm B} > -15$) show a larger scatter in the specific star formation rate as compared to bright galaxies. They suggest that this could be because the processes that regulate star formation in bright spirals are not operative in dwarfs, consistent with the finding of \cite{bc06} of a lack of a simple relationship between gas column density and star formation rate in dwarf galaxies. Fig.~\ref{fig:itf} indicates that even integration over time is not sufficient to remove this scatter, i.e. even the integrated star formation rate of dwarf galaxies shows substantially more scatter than that of bright spirals. In a recent study using a sample of SDSS galaxies, \cite{geha06} found a similar increase in scatter around the I band TF relation at low luminosities. However, in contrast to what we find here they do not find that faint galaxies are systematically under-luminous. The reason for this discrepancy is unclear, though we do note that we are using galaxies with accurately measured distances, while \cite{geha06} used Hubble flow based distances for their galaxies.

%

\begin{figure*}
\psfig{file=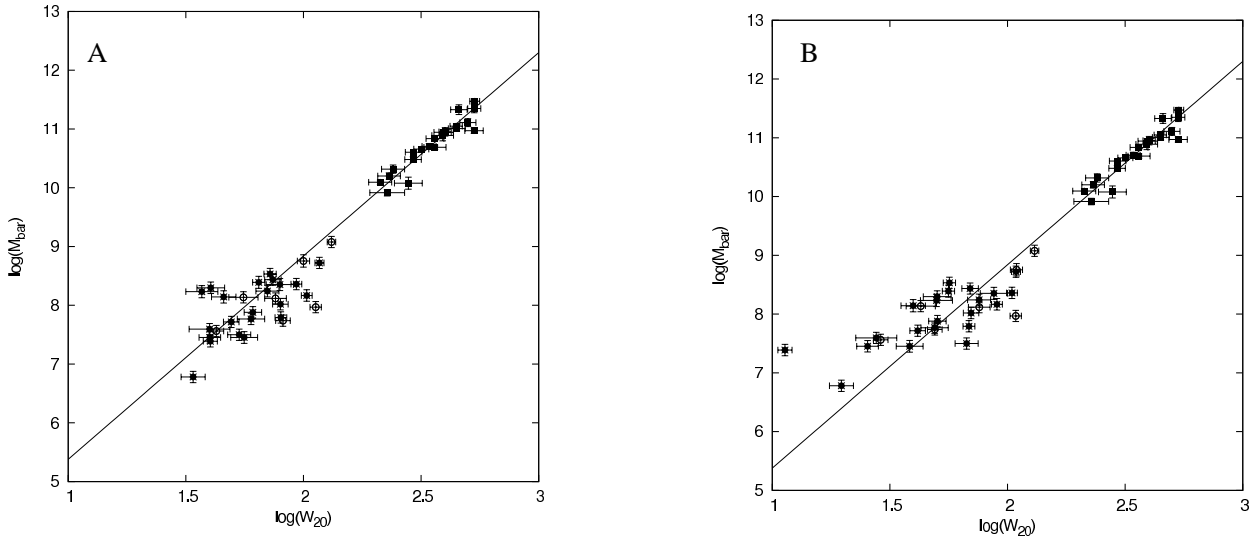,width=6.5truein}
\caption{The solid lines show the BTF for the bright HST sample,
Panel[A] is for a Salpeter ``lite'' IMF, while Panel [B] is for a Botemma IMF. The hollow circles are for FIGGS galaxies without TRGB distances, the filled circles are FIGGS galaxies with TRGB distances and the filled square are galaxies from the HST sample. }
\label{fig:btf}
\end{figure*}

\begin{figure}
\psfig{file=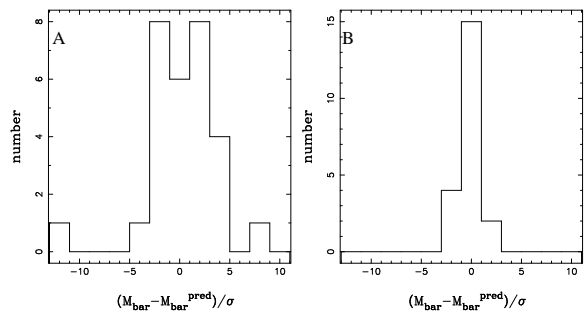,width=3.0truein}
\caption{Scatter about the combined best fit BTF for the FIGGS (Panel [A]) and HST (Panel [B]) galaxies. The BTF used to generate this figure is that between the Baryonic mass derived from the Salpeter ``lite'' IMF and log(W$_{20}$). Different choices of IMF or circular velocity indicator do not qualitatively affect the plot. The scatter has been scaled by the combined measurement errors.}
\label{fig:btf_dev}
\end{figure}

As mentioned in the introduction, the deviation of dwarf galaxies
from the B band TF relation was one of the primary motivations for searching for a relation between the total baryonic mass and the
velocity width, i.e. the baryonic Tully Fisher (BTF) relation 
(\cite{mg00}). The BTF relation for the HST sample is shown in
Fig.~\ref{fig:btf} as a solid line. As can be seen, the FIGGS sample deviates much less from this relation that it does from the I band TF relation. The average deviation of the FIGGS galaxies from the BTF relation set by bright spirals is -0.16(0.02),-0.11(0.02) for the Salpeter ``lite'' and Bottema IMFs and using V$_{rot}$ as the circular velocity indicator, and 0.09(0.02), 0.09(0.02) for the same two IMFs but with W$_{20}$ as the circular velocity indicator. Clearly, whether one sees a deviation from the BTF at the faint end seems dominated by  systematic uncertainties (IMF, choice of circular velocity indicator). On the other hand, the increased scatter about the BTF at the faint mass end is independent of the choice of IMF or circular velocity indicator.  Fig.\ref{fig:btf_dev}) compares the scatter about the combined best fit BTF for the HST and FIGGS galaxies, the difference
in scatter is striking. Different choices of IMF or circular velocity indicator do not qualitatively change the plot.

\begin{figure}
\psfig{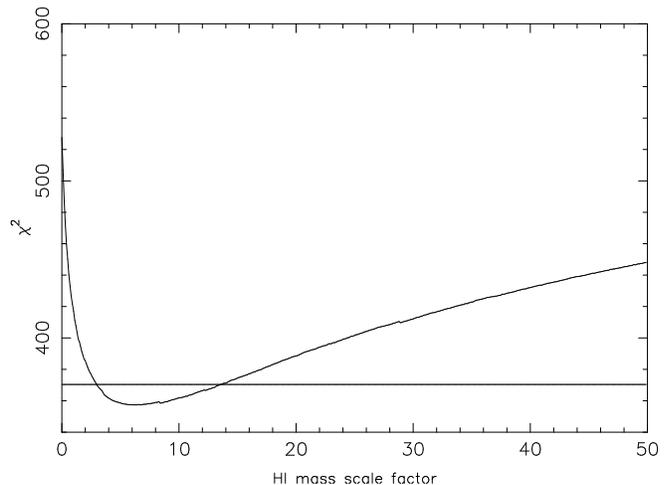}
\caption{Models with HI mass scale factor as a free parameter. The graph shows $\chi^2$ as a function of the HI mass scale factor. The minimum $\chi^2$ is for a mass scale factor $\sim 6.7$. The horizontal line shows the threshold $\chi^2$ for a 95\% significance that the model with a variable HI mass scale factor provides a better fit to the data.}
\label{fig:mhi_scale}
\end{figure}

\begin{figure}
\psfig{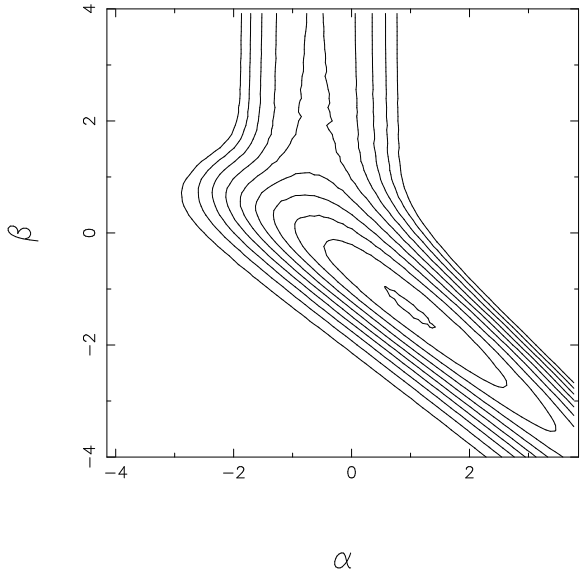}
\caption{Contours of $\chi^2$ as a function of the stellar mass to light ratio, $log(\Gamma_{*}) = \alpha + \beta (B-V)$. The scale factor for the HI mass has been kept fixed at 1.33. The contour
levels start at 358 and are uniformly spaced in steps of 20.}
\label{fig:imf_scale}
\end{figure}

As discussed above, systematic uncertainties limit the measurement of the BTF. Numerous authors (e.g. \cite{pfenniger05,mg05} have tried to approach this problem along the reverse direction, viz. to use the BTF itself to determine one or both of the scaling of the HI mass to the total gas mass and the stellar mass to light ratio. 
In this approach, one assumes that an intrinsic BTF relation exists, and that the right choice of 
stellar mass to light ratio and/or HI mass scaling will minimize the scatter of the observed 
data about the relationship. Pfenniger \& Revaz (2005)  found that the BTF
relation is optimally improved when the HI mass is multiplied
by a factor of 3; however the actual best fit value is poorly
constrained, and scale factors of up to $\sim 11$ lead to a
tightening of the BTF. Allowing a free scaling of the HI mass is
prompted by models  which have a dark baryonic component whose
distribution is correlated with that of HI (e.g Hoekstra et al. 2001).
These models inturn are based on the possibility
that some of the dark matter in the halos of galaxies may be in the
form of cold,dense gas clouds (Gerhard \& Silk (1996);
Pfenniger \& Combes (1994)). Many observations based on
dynamical and stability analysis indicate that galactic disks may be
more massive than inferred from its luminous mass (Fuchs, B. (2003)
; Masset \& Bureau (2003); Pfenniger \& Revaz (2005)).
Further, from theoretical modelling and N-body
simulations, Pfenniger \& Revaz (2004) found that
if disks are marginally stable with respect to bending instabilities,
the mass within the HI disks must be a multiple of that detected
in HI and stars.
We note that since the stellar mass and HI mass are highly 
correlated (correlation coefficient $\sim 0.91$ between log(M$_{\rm star}$) and log(M$_{\rm HI}$), simultaneous minimization of both the stellar mass to light ratio and the scale factor for the HI mass is degenerate. So instead we first allow the HI mass scale factor to vary keeping the mass to light ratio of the stellar disk fixed (Fig.~\ref{fig:mhi_scale}) and then allow the stellar mass to light ratio to vary keeping the HI mass scale factor fixed (Fig.~\ref{fig:imf_scale}). In the case of a variable HI mass scale factor (Fig.~\ref{fig:mhi_scale}), there is a broad minimum in the $\chi^2$, around a scale factor of $\sim 6.7$. The minimum $\chi^2$ is 357.2, as compared to 402.3 for the case in which the HI mass scale factor is kept at 1.33. Since the number of parameters has been increased, a decrease in the $\chi^2$ is to be expected. The significance of the decrease can be estimated using the F-test. For
nested models with $\nu{_1}$ and $\nu_{2}$ degrees of freedom respectively ($p=\nu{_1}-\nu{_2}$) being the extra number of parameters in the second model) and $\chi^2$ values $\chi{_1}^2$  and $\chi{_2}^2$ respectively, the F-test looks at the value 

$$ F = {{(\chi{_1}^2 - \chi{_2}^2)/p} \over \chi{_2}^2/\nu_{2}} $$

Large values of F indicate a statistically significant improvement in the fit. Applying the F-test to the $\chi^2$ numbers above indicates that this decrease is significant at the $\sim 2\%$ level, i.e. there is a $2\%$ probability that a HI scale factor of 6.7 does not provide a better fit to the data. The horizontal line in Fig.~\ref{fig:mhi_scale} indicates the threshold value of $\chi^2$ for a $5\%$ significance. As can be seen this leads to a relatively broad range in the scale factor, i.e. from $\sim 3.0 - 13.5$. If one excludes the FIGGS galaxies without TRGB distances, the minimum $\chi^2$ occurs at an HI mass scale factor of $\sim 5$ and has a 
$\sim 5\%$ significance.

      For models with a fixed HI mass scale factor of 1.33 but variable stellar mass to light ratio, we follow the parametrization of \cite{bell03} and assume that the stellar mass to light ratio is given by log$(\Gamma_{*}) = \alpha + \beta (B-V)$, where $\alpha$ and $\beta$ are free parameters. Fig.~\ref{fig:imf_scale} shows contours of the resulting $\chi^2$. The minimum $\chi^2$ (356.3) corresponds to $\alpha = -1.51, \beta = 1.20$. For comparison, $\alpha = -0.549, \beta = 0.824$ corresponds to the Salpeter ``lite'' IMF and $\alpha = -0.749, \beta = 0.824$ corresponds to the Bottema IMF. The decrease in the $\chi^2$ however is significant only at the $\sim 7\%$ level (as compared to the $\sim 2\%$ level for a variable HI mass scale factor).

\begin{figure}
\psfig{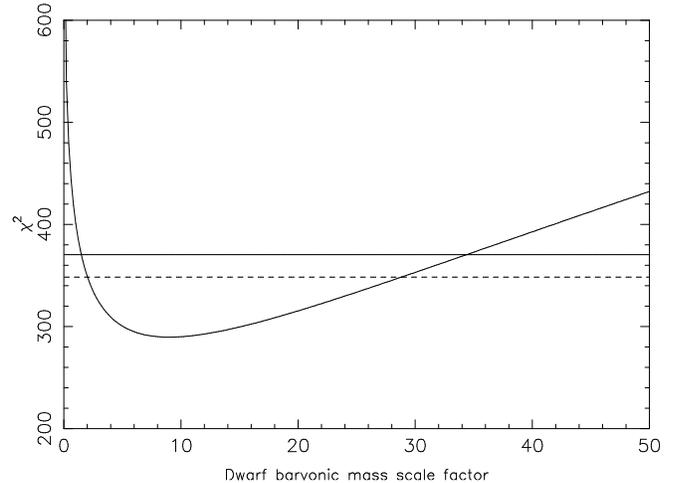}
\caption{Models in which the scale factor of the baryonic mass of the galaxies in the FIGGS sample is a free parameter. The graph shows $\chi^2$ as a function of the baryonic mass scale factor. The minimum $\chi^2$ is for a mass scale factor $\sim 9.2$. The solid horizontal line shows the threshold $\chi^2$ for a 95\% significance that the model with a variable HI mass scale factor provides a better fit to the data.The dashed horizontal line is the threshold $\chi^2$ for a 99\% significance.}
\label{fig:mbar_scale}
\end{figure}

\begin{figure}
\psfig{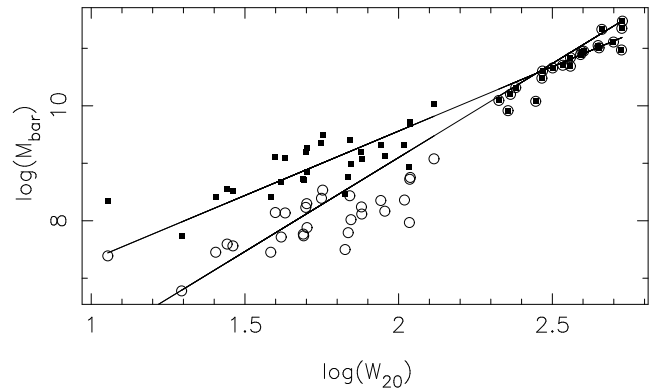}
\caption{Baryonic Tully Fisher relations. The hollow points are for the FIGGS galaxies with stellar mass to light ratio computed using a Salpeter ``lite'' IMF. For the solid points the total baryonic mass has been increased by a factor of 9.2. For the HST galaxies (the cluster of points at large log(W$_{20}$)) the baryonic mass is not scaled. The two lines show the best fit BTF for these two cases.}
\label{fig:vmbar_btf}
\end{figure}

    Since the HI mass forms a much larger fraction of the total baryonic mass in the case of galaxies from the FIGGS sample, the greater sensitivity of the goodness of fit to the HI mass scale factor suggests that models in which one has a scale factor for the total baryonic mass of dwarf galaxies would probably significantly improve the fit. While this (like the earlier two models) is a phenomenological model, it would correspond to galaxy formation models in which dwarf galaxies have a larger fraction of ``dark'' (e.g. highly ionized) or missing baryons. For e.g. models which account for the reheating and ionization of the gas during the epoch of recombination predict that large galaxies should have essentially the 
cosmic baryon fraction, while small galaxies (with circular velocity
in the $\sim 20$~km/s range) could lose up to 90\% of their baryons.
Of course, in such a situation, when applying the BTF relation with the observed velocity widths, we have to make the further assumption that the baryon loss does not affect the observed velocity width. The $\chi^2$ curve for a model in which the scale factor of the baryonic mass of the FIGGS galaxies is a free parameter (but the scale factor for the baryonic mass of the HST sample is kept at 1.0) is shown in Fig.~\ref{fig:mbar_scale}. Once again, the $\chi^2$ curve shows a broad minimum, however the minimum $\chi^2$ (289.6, for a scale factor of 9.18) is substantially lower than in the case that only the HI mass was rescaled. The F-test gives a probability of $\sim 2\times 10^{-4}$ that this reduction in $\chi^2$ is by chance.  The horizontal line in Fig.~\ref{fig:mbar_scale} indicates the threshold value of $\chi^2$ for a $1\%$ significance. As can be seen this leads to a relatively broad range in the scale factor, i.e. from $\sim 2.0 - 28.8$.
If one includes only those FIGGS galaxies with
TRGB distances, the minimum $\chi^2$ occurs at a mass scale factor
of $\sim 7.56$ and corresponds to a significance of $\sim 10^{-3}$.
The best fit ``BTF'' relation after the baryonic mass of the FIGGS galaxies has been scaled by a factor of 9.18 is shown in Fig.~\ref{fig:vmbar_btf}. For reference, the original BTF relation without this rescaling is also shown. Rescaling by a factor of 9.18 implies that over 90\% of the original baryonic mass is ``dark'' or ``missing''.

To summarize, we present TF relations for a sample combining extremely faint dwarf irregular galaxies from the FIGGS survey and large spiral galaxies. For the FIGGS sample alone, we find that though the median HI mass and stellar mass are comparable, the HI mass correlates with the velocity width much better than the stellar mass. Distances are accurately known for the bulk of the galaxies in our sample, and we find that the extremely faint dwarfs show substantial intrinsic scatter about both the I band TF and the BTF relations. Deviations from the BTF at the faint end are sensitively dependent on systematics, e.g. the assumed stellar mass to light ratio. If one assumes that there is an intrinsic baryonic TF relation one could try to determine the total baryonic mass by searching for the prescription that leads to the tightest BTF. We find that changing the scale factor of the total HI mass leads to a more significant tightening of the BTF than changing the stellar mass to light ratio. Finally, the most significant tightening that we find is when we allow the entire baryonic mass of the faint (but not the bright) galaxies to scale by
a constant factor. This best fit relation is for a scale factor $\sim 9$. The exact value is however poorly constrained, any value between $\sim 2 - 28.8$ leads to a significant tightening of the BTF.

\section*{Acknowledgments}

        The observations presented in this paper were made
with the Giant Metrewave Radio Telescope (GMRT). The GMRT is operated
by the National Center for Radio Astrophysics of the Tata Institute
of Fundamental Research.


\begin{thebibliography}{}
\bibitem[\protect\citeauthoryear{Begum et al.}{2008}]{begum08} Begum, A.,   Chengalur, J.N.,
	Karachentsev I. D., Sharina, M. \& Serafim, S. 2008, MNRAS submitted.
\bibitem[\protect\citeauthoryear{Begum et al.}{2006}]{bc06} Begum, A.,   Chengalur, J.N., Karachentsev I. D., Sharina, M. \& Serafim, S. 2006, MNRAS, 365, 1220. 
\bibitem[\protect\citeauthoryear{Begum et al.}{2005}]{bc05} Begum, A.,   Chengalur, J.N. \&
	Karachentsev I. D.  2005, A\&A, 433, L1 
\bibitem[\protect\citeauthoryear{Begum et al.}{2003}]{b1} Begum, A.,  
	Chengalur, J.N. \& Hopp, U., 2003, New Astronomy, 8, 267
\bibitem[\protect\citeauthoryear{Begum \& Chengalur}{2004}]{bc04} Begum, A \& 
	Chengalur, J.N., 2004, A\&A, 424, 509
\bibitem[\protect\citeauthoryear{Bell et al.}{2003}]{bell03} Bell, E. F., McIntosh, D. H., Katz, N.
	\& Weinberg, M. D., 2003, ApJSS, 149, 289
\bibitem[\protect\citeauthoryear{Cote et al.}{2000}]{cote00} Cote, S., Carignan, C. \& Freeman, K. C., 2000, AJ, 120, 3027
\bibitem[\protect\citeauthoryear{Dekel \& Woo}{2003}]{dw03} Dekel, A. \& Woo, J., 2003, MNRAS, 344, 1131
\bibitem[\protect\citeauthoryear{Efstathiou}{2000}]{ef00} Efstathiou, G.  2000, MNRAS, 317, 697
\bibitem[\protect\citeauthoryear{Fuchs}{2003}]{fuch03} Fuchs, B. 2003, Ap\&SS, 284, 719
\bibitem[\protect\citeauthoryear{Geha et al.}{2006}]{geha06} Geha, M., Blanton, M. R., Masjedi, M. \& 
	West, A. A., 2006, ApJ, 653, 240
\bibitem[\protect\citeauthoryear{Gerhard \& Silk}{1996}]{gerhard96} Gerhard, O. \& Silk, J., 1996, ApJ, 472, 34
\bibitem[\protect\citeauthoryear{Gnedin}{2000}]{gnedin00} Gnedin N.~Y., 2000, ApJ, 542, 535 
\bibitem[\protect\citeauthoryear{Hoekstra et al.}{2001}]{hk01}Hoekstra, H., van Albada, T. S.
        \& Sancisi, R., 2001, MNRAS, 323, 453
\bibitem[\protect\citeauthoryear{Holtzman et al.}{1995}]{holtz95} Holtzman, J. A., Burrows, C. J.,
	Casertano, S., Hester, J. J., Trauger, J. T., Watson, A. M. \& Worthey, G., 1995, PASP, 107, 1065
\bibitem[\protect\citeauthoryear{Jordi et al.}{2006}]{jordi06} Jordi, K., Grebel, E. K. \& Ammon, K., 2006, A\&A, 460, 339
\bibitem[\protect\citeauthoryear{Karachentsev et al.}{2004}]{kk04} Karachentsev I. D., Karachentseva V. E., Huchtmeier W. K.
	\& Makarov D. I., 2004, AJ, 127, 2031
\bibitem[\protect\citeauthoryear{Lee et al.}{1999}]{lee99} 
Lee M.~G., Aparicio A., Tikonov N., Byun Y.-I., Kim E., 1999, AJ, 118, 853
\bibitem[\protect\citeauthoryear{Lee et al.}{2007}]{lee07} 
Lee J.~C., Kennicutt R.~C., Funes J.~G., J.~S., Sakai S., Akiyama S., 2007, 
arXiv, 711, arXiv:0711.1390
\bibitem[\protect\citeauthoryear{Lo et al.}{1993}]{lo93} Lo, K. Y., Sargent, W. L. W. \& Young, K., 1993, AJ, 106, 507
\bibitem[\protect\citeauthoryear{Makarova}{1999}]{makarova99} Makarova L., 1999, A\&AS, 139, 491 
\bibitem[\protect\citeauthoryear{Masset & Bureau}{2003}]{masset03} Masset, F. S. \& Bureau, M., 2003, ApJ, 596, 152
\bibitem[\protect\citeauthoryear{McGaugh, S.}{2005}]{mg05} McGaugh, S. S., 2005, ApJ, 632, 859
\bibitem[\protect\citeauthoryear{McGaugh et al.}{2000}]{mg00} McGaugh, S. S., Schombert, J. M., Bothun, G. D. \&
	de Blok, W. J. G., 2000, ApJ, 533, 99
\bibitem[\protect\citeauthoryear{Ott et al.}{2001}]{ott01} Ott, J., Walter, F., Brinks, E., Van Dyk, S. D.,
        Dirsch, B. \& Klein, U., 2001, AJ, 122, 3070
\bibitem[\protect\citeauthoryear{Pfenniger \& Combes}{1994}]{pfenniger94}Pfenniger D. \& Combes F., 1994, A\&A, 285, 94.
\bibitem[\protect\citeauthoryear{Pfenniger \& Revaz}{2004}]{pfenniger04} Pfenniger D. \& Revaz Y., 2004, ASSL, 319, 307.
\bibitem[\protect\citeauthoryear{Pfenniger \& Revaz}{2005}]{pfenniger05} Pfenniger D. \& Revaz Y., 2005, A\&A, 431, 511.
\bibitem[\protect\citeauthoryear{Sakai et al.}{2000}]{sakai00} Sakai, A., Jeremy, M. R. et al., 2000, ApJ, 529, 698
\bibitem[\protect\citeauthoryear{Sharina et al.}{2007}]{sme07} Sharina M.~E., et al., 2007, arXiv, 712, arXiv:0712.1226 
\bibitem[\protect\citeauthoryear{Schombert et al.}{1997}]{sc97} Schombert, J. M., Pildis, R. A. \&
         Eder, J. A., 1997, ApJS, 111, 233
\bibitem[\protect\citeauthoryear{Sirianni et al.}{2005}]{se05} Sirianni, M., et al., 2005, PASP, 117, 1049
\bibitem[\protect\citeauthoryear{Tassis et al.}{2007}]{ts07} Tassis, K., Kravtsov, A. V. \& Gnedin, N. Y.,
	2007, astro-ph/0609763v2
\bibitem[\protect\citeauthoryear{Taylor, Kobulnicky \& Skillman}{1998}]{taylor98} Taylor, C. L., Kobulnicky, H. A. \& Skillman, E. D. 1998, A.J.,116,2746-2756.
\bibitem[\protect\citeauthoryear{Verheijen}{2001}]{verheijen01} 
Verheijen M.~A.~W., 2001, ApJ, 563, 694 
\bibitem[\protect\citeauthoryear{Verheijen \& Sancisi}{2001}]{vs01} Verheijen, M. A. W. \& Sancisi, R., 2001, A\&A, 370, 765
\end{thebibliography}
\end{document}